\documentclass[apjl]{emulateapj}

\shorttitle{Warm Gas at High $z$}
\shortauthors{E.~R.~Stanway et al.}

\begin{document}

\title{Large Scale Structure traced by Molecular Gas at High Redshift}
\author{Elizabeth R.~Stanway, Malcolm N.~Bremer, Luke J.~M.~Davies, Mark Birkinshaw}
\affil{H H Wills Physics Laboratory, Tyndall Avenue, Bristol, BS8 1TL, UK}
\and
\author{Laura S.~Douglas, Matthew D.~Lehnert}
\affil{Laboratoire d'Etudes des Galaxies, Etoiles, Physique et Instrumentation GEPI, Observatoire de Paris, Meudon, France}

\begin{abstract}

  We present observations of redshifted CO(1-0) and CO(2-1) in a field
  containing an overdensity of Lyman break galaxies (LBGs) at
  $z=5.12$. Our Australia Telescope Compact Array observations were
  centered between two spectroscopically-confirmed $z=5.12$
  galaxies. We place upper limits on the molecular gas masses in these
  two galaxies of M(H$_2$)$<1.7 \times 10^{10}$ M$_\odot$ and
  $<2.9 \times 10^{9}$ M$_\odot$ (2\,$\sigma$),
  comparable to their stellar masses.  We detect an
  optically-faint line emitter situated between the two LBGs which we
  identify as warm molecular gas at $z=5.1245\pm0.0001$. This source,
  detected in the CO(2-1) transition but undetected in CO(1-0), has an
  integrated line flux of $0.106 \pm 0.012$ Jy km\,s$^{-1}$, yielding
  an inferred gas mass M(H$_2$)=$(1.9\pm0.2)\times
  10^{10}$\,M$_\odot$.  Molecular line emitters without detectable
  counterparts at optical and infrared wavelengths may be
  crucial tracers of structure and mass at high redshift.

\end{abstract}


\section{Introduction}
\label{sec:intro}

Lyman-break galaxies have given us our first detailed look at star
formation at high redshifts ($z>4$).  However, our picture
of the distant universe is limited because these objects are selected
for strong rest-frame ultraviolet emission arising from
substantial unobscured starbursts in the galaxies. Individually, they
tell us little about darker baryons at the same redshift, whether the
systems containing the baryons are intrinsically less UV-luminous or
are obscured by dust.

The Lyman-break galaxy (LBG) population has now been probed from $z=2$
\citep{2004ApJ...604..534S} up to $z=6$ \citep{2007MNRAS.376..727S},
with tentative detections reported at redshifts as high as $z=10$
\citep{2005ApJ...624L...5B}. The photometric and spectroscopic
properties of LBGs evolve with increasing redshift. Typical
LBGs at $z\approx5$ have stellar masses of a few times
10$^9$\,M$_\odot$, are predominantly young ($<50$\,Myr), and are
sub-solar but not primordial in their metallicity
(Z$\approx$0.2\,Z$_\odot$), where these properties are deduced from
their rest-frame optical and near-infrared spectral
energy distribution \citep{2007astro.ph..1725V}.

However something is missing in our picture of the
universe at $z>5$. The youth of Lyman-break objects when compared to
the range of lookback times probed in a given survey suggests that
only a small fraction of the overall population is undergoing a phase
of strong unobscured star formation, and hence is detectable in the
rest UV, at a given time. Fossil stellar populations in the most
massive galaxies at the present time suggest that the bulk of their
stars were formed at $z>3$ \citep{2004Natur.428..625H}, but the mass
assembly history of the universe as traced by LBGs implies that only
1\% of the baryonic material that will eventually form today's
galaxies was in collapsed systems at $z=5$. Where is the other 99\%?
The bulk of the baryonic material at these high redshifts must be in
either low mass systems or the gas phase.

Damped Lyman-alpha systems in the spectra of distant quasars have
shown that dense neutral gas is ubiquitous at high redshift,
holding a third of the HI atoms at $z=5$ and 
appearing in 10\% of sight-lines per unit cosmological distance
\citep{2005ApJ...635..123P}. These sources represent reservoirs of
neutral gas for star formation, yet are known to be deficient in H$_2$
relative to local galaxies. Molecular gas is instead more tightly
concentrated in high-density, relatively dusty regions
\citep{2006ApJ...643..675Z}.

Securely identifying this material is challenging.  At $z\approx5$,
the far-infrared lines emitted by cool and obscured material are
redshifted to millimeter wavelengths, but not necessarily into an
atmospheric window. At these extreme luminosity distances lines
are faint, requiring long integration times to secure even a
tentative detection. The fields of view of existing millimeter
interferometers are small, and their
correlators have, until recently, been limited to bandwidths of a
few hundred MHz (equivalent to $\Delta z/z < 0.01$). The resulting
small volumes made survey work at high redshift unfeasible. As a
result, the few detections of line emission from molecular and atomic
gas at high redshift have been towards active galactic nuclei -
quasars to $z=6.4$ \citep{2003A&A...409L..47B}, radio galaxies to
$z=5.2$ \citep{2005ApJ...621L...1K} - and hence probed highly atypical
environments.

In this paper we present the results of a pilot programme to explore
the cool gas associated with a large scale structure at $z>5$ marked
out by Lyman-break galaxies rather than by an active galaxy. Our 40
arcmin$^2$ target field contains seven UV-luminous sources with
spectroscopically confirmed redshifts in a narrow range ($\Delta z <
0.1$), a 6$\sigma$ excess over the typical density of such sources in
our survey, but shows no clear spatial clustering\footnote{The
detailed properties of this field, and of others probed in our ESO
Remote Galaxy Survey (ERGS) will be discussed in a forthcoming paper
by Douglas et al.}. Given the redshift dispersion and spatial
distribution of these galaxies, it is clear that their star formation
cannot be a response to a common triggering event, and yet the
probability of such a configuration arising by chance is small. The
simplest explanation is that there is a large mass of hidden baryons
in this field at this redshift. At any one time this structure is
traced by the small fraction of that material undergoing comparatively
short-term UV-luminous star formation events and appearing as LBGs.
While the stars within the LBGs are likely to end up in the most
massive current-day galaxies after subsequent mergers, the total
stellar mass in even this overdensity of LBGs is an order of magnitude
too small to account for the stars in the most massive
spheroids. Consequently, there should be far more baryons in the
vicinity of the LBGs than revealed in the UV. Our pilot programme was
designed to probe the cool gas mass in a small section of the
structure, centered on a pair of LBGs with Lyman-$\alpha$ emission
redshifts separated by $\Delta z=0.004$. The observations target not
only the two LBGs, but any darker/fainter systems in the underlying
large scale structure.  The results for lower redshift LBGs
\citep[e.g.][]{2004ApJ...604..125B}, suggested that we were unlikely
to detect the two known galaxies directly.

All magnitudes in this paper are quoted in the AB system
\citep{1983ApJ...266..713O}.  We adopt a $\Lambda$CDM cosmology with
($\Omega_{\Lambda}$, $\Omega_{M}$, $h$)=(0.7, 0.3, 0.7).

\section{Observations at the ATCA}
\label{sec:obs}

The observations were carried out at the Australia Telescope Compact
Array in March-April 2008. The array was in the relatively compact
H168 configuration, with both North-South and East-West baselines. We
used the correlator in the 128 MHz/64 channel configuration with two
overlapping intermediate frequency (IF) bands, giving a total
bandwidth of 240MHz, and measuring the total incident flux density,
but not polarization.

We investigated two transitions of CO at high redshift. In order to
constrain the CO(1-0) line, we tuned the IF bands to 18.838\,GHz and
18.958\,GHz (allowing detection of an emission line in the
range $z=5.06-5.14$). In this configuration, the spectral resolution
is 4\,MHz or 63.5\,km\,s$^{-1}$. Observations were taken in three 8\,hr
observing periods on the nights of 2008 March 22-24
(although data taken on the 23rd were discarded due to poor
weather). A single pointing was observed for the entire observing
period, and a nearby bright source (1045-188) observed every 15
mins to determine phase stability. Pointing accuracy was checked every
hour.  Absolute flux calibration was through observations of 1921-293
(the standard flux calibrator at the ATCA) each night.  The
half-power-beam-width (HPBW) of the ATCA at this frequency is 2.5
arcmin and the naturally-weighted restoring beam in this configuration
is 15.9\arcsec $\times$ 10.9\arcsec.

The CO(2-1) line was observed in four 11\,hr observations on the
nights of 2008 Apr 04-07 using the recently-commissioned 7mm receiver
system. The IF bands were tuned to 37.672\,GHz and 37.792\,GHz (giving
redshift coverage $z=5.09-5.13$). A single pointing
centered on the 18.8\,GHz field center was observed, and 1045-188 was
used as a pointing and phase calibrator. Uranus, the primary flux
calibrator for the ATCA at millimeter wavelengths, was not visible at
any point in our observing period. Instead we bootstrap our primary
flux calibration from two compact HII regions, observed each night,
which were also observed using our configuration by ATCA staff on the
afternoon of 2008 Apr 04 together with Uranus and a bandpass
calibrator. 
At this frequency, the HPBW is 74\arcsec\ and the
restoring beam is 7.3\arcsec $\times$ 4.8\arcsec.

The total useful observing time was 16 hours at 18.9\,GHz and 44 hours at
37.7\,GHz. The RMS noise in the final images is 0.11 mJy/beam and 0.10
mJy/beam in each 4\,MHz channel at 18.9 and 37.7\,GHz
respectively.

\begin{table*}
\begin{center}
\caption{High-$z$ sources in our target field.}
\begin{tabular}{clcccc}
Source & RA \& Declination (J2000) & z     & $I_{AB}$       & $S_{CO}\Delta v$  & M(H$_2$) \\
\tableline\tableline
A & 10:40:46.66 -11:58:33.8        & 5.120 & 27.00$\pm$0.57 & -                 & $<1.7$     \\
B & 10:40:41.51 -11:58:21.6        & 5.116 & 25.71$\pm$0.21 & -                 & $<0.29$    \\
C & 10:40:43.50 -11:57:56.8        & 5.125 & $>26.63$       & $0.106 \pm 0.012$ & $1.9\pm0.2$\\
\end{tabular}
\tablecomments{Sources A and B
are spectroscopically confirmed Lyman-break galaxies, while source C
is the millimeter line-emitter discussed in
section~\ref{sec:co_detect}. $S_{CO}\Delta v$ is in units of Jy km
s$^{-1}$ and M(H$_2$) in units of $10^{10}$ M$_\odot$ as discussed in
the text. 2\,$\sigma$ limits are given where appropriate.}\label{tab:sources}
\end{center}
\end{table*}

\section{Limits on the CO in Lyman Break Galaxies}
\label{sec:lbg_limits}

\begin{figure}
\plotone{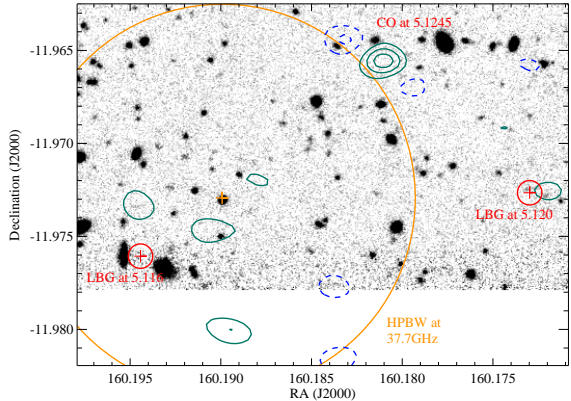}
\caption{Contour map of 6-channel multi-frequency synthesis image
  centered at 37.64\,GHz, overlayed on an $I$-band image of the same
  region. The positions of the two Lyman break galaxies at $z=5.120$
  and $z=5.116$ respectively are shown as small circles. The half
  power point of the synthesized beam at 37.7\,GHz is indicated by a
  large circle. The HPBW at 18.9\,GHz is twice as large and
  encompasses the whole image.  Contours mark formal 2.5, 5 and 7.5\,$\sigma$
  surface brightness levels. Negative contours are shown with dashed lines. Although the
  right-hand LBG is situated within a few arcseconds of a 2.5\,$\sigma$
  contour, this is part of an extended low-level artifact. \label{fig:image}}
\end{figure}

As illustrated in figure \ref{fig:image}, our pointing included
two Lyman-$\alpha$ emitting, Lyman-break galaxies within the
HPBW at 18.9\,GHz (table \ref{tab:sources}). At 37.7\,GHz one of these
(at $z=5.116$) lies well outside the beam half-power point, while the
second (at $z=5.120$) remains close to the pointing center.  Neither
galaxy shows evidence for line or continuum emission in either of the
two transitions surveyed.  This is not, in itself, surprising given
the youth and low masses \citep[typically a few $\times 10^9$\,M$_\odot$,
][]{2007astro.ph..1725V} of $z=5$ Lyman-break
galaxies. However, these observations provide the strongest
constraint yet on the molecular gas content of $z>5$ star-forming
galaxies.

We use our limiting flux at the frequency of the redshifted CO(1-0)
line to constrain the molecular gas content of these galaxies. We
employ the widely-used conversion factor of
M$_{\mathrm{gas}}$/$L'_\mathrm{CO}$=0.8 M$_\odot$
(K\,km\,s$^{-1}$\,pc$^2$)$^{-1}$ based on local infrared-luminous
galaxies \citep{2005ARA&A..43..677S}. Our image RMS noise of 0.11
mJy/beam at the CO(1-0) line corresponds to a formal 2 sigma limit
$L'_\mathrm{CO}<2.2 \times 10^{10}$ K\,km\,s$^{-1}$\,pc$^{2}$ for the
luminosity of an unresolved 200 km s$^{-1}$ emission line at $z=5.12$.
Thus we constrain the molecular gas mass of the known Lyman-break
galaxies in this field to be M(H$_2$)$<1.7 \times 10^{10}$\,M$_\odot$
(2$\sigma$).  

One of our LBGs (source B in table \ref{tab:sources}) is also close to
the center of the beam in our 37\,GHz observations.  For thermalized,
optically-thick CO emission, the line luminosity is independent of
transition. Making this assumption, non-detection of the 2-1 line
places a tighter constraint on the mass of M(H$_2$)$<2.9 \times
10^{9}$\,M$_\odot$ (2$\sigma$), comparable to the inferred stellar
mass in such sources.

\section{An Optically-faint Molecular Line Emitter}
\label{sec:co_detect}

Inspection of the ATCA data cubes revealed no evidence for continuum
sources. Only one emission feature was found to extend over more than
two adjacent frequency channels. This feature, a spatially-unresolved
emission line centered at 37.642\,GHz with a FWHM of 110\,km\,s$^{-1}$
(equivalent to four independent frequency channels), lies on the beam
half-power point.  It has a peak flux of $0.94 \pm 0.15$\,mJy/beam, and
a velocity integrated flux of $0.106 \pm 0.012$ Jy km s$^{-1}$.  A
combined frame of channels not containing line emission (a total
bandwidth of 180\,MHz) shows no significant continuum
emission with a formal limit on the continuum of $<$0.49 mJy/beam
at 37.73\,GHz. The line emission is centered 1 arcminute (370\,kpc,
comoving) north-west and 40 arcseconds (250\,kpc) north-east of our
spectroscopically confirmed Lyman-break galaxies at $z=5.120$ and
$z=5.116$ respectively, towards more LBGs in the
field.

This line emitter is undetected at 18.9\,GHz and not spatially
coincident with a known source at any observed wavelength (from
0.5$\mu$m to 21cm). Our optical and IR data require that
the host galaxy must have a flux density $<0.2\,\mu$Jy at an observed
frame wavelength of 4.5$\mu$m (0.7$\mu$m rest if at $z=5.125$) and
$<0.08\,\mu$Jy at a 0.8$\mu$m (1300\AA\ rest). No sub-mm data exist at
this location. We note that the lensing properties of this field are
well understood, and we use the lensing maps of
\citet{2006A&A...451..395C} to estimate a maximum 10\% 
enhancement in the flux of a source at this location.

In order to verify the reality of the 37.6\,GHz line, we investigated
its properties thoroughly. The velocity-integrated line is a
formal 8\,$\sigma$ detection in a multi-frequency synthesis image of 6
channels centered on the line emission. It is detected in
the $uv$-dataset on multiple baselines, and in two independent
sub-sets of the data. Its strength in the flux-calibrated image is
independent of image deconvolution algorithm and the location of the
imaging center. 
We conclude that this is unlikely to be a spurious detection.

We evaluate the probability that this is a line emitter at a redshift
unrelated to our high redshift system (i.e. not a CO(2-1) emission
line), by considering the population models of
\citet{2000MNRAS.313..559B}.  These authors calculated the expected
number density of line emitters as a function of frequency and line
flux, given models for galaxy and dark matter distributions at
$z<10$. As a cautious assumption, we consider the source counts
predicted at 50\,GHz at our integrated flux density (while noting that
source counts at 37.6\,GHz are likely to be lower).  In our
256\,MHz total bandwidth, and allowing for source detections within
1.5 HPBW, we would expect 0.003 line emitters in our
survey, with these dominated by redshifted CO(1-0) lines at $z=2.1$.
Coupled with the known existence
of structure at $z=5.12$ in this field, we
interpret the line detection as CO(2-1) emission at
$z=5.1245\pm0.0001$ at high confidence.

At the measured line flux in the CO(2-1) transition, and assuming
constant luminosity, the expected counterpart in the 1-0 transition
would be below the 2\,$\sigma$ level at 18.8\,GHz, consistent with our
non-detection. Using the same CO-to-H$_2$ conversion factor discussed
in section \ref{sec:lbg_limits} applied to the luminosity of the
CO(2-1) transition, we infer a molecular gas mass of
M(H$_2$)$=(1.9\pm0.2) \times 10^{10}$ M$_\odot$ in this source.

The beam width at 37.7\,GHz implies an upper limit on the
physical size of our unresolved source of 45$\times$30\,kpc. A simple
virial calculation assuming a uniform mass distribution suggests that
our derived gas mass could represent up to 50\% of the dynamical mass
in the system, while remaining both bound and smaller than our beam
size.

\section{Large Scale Structure at $z=5.12$}
\label{sec:discuss}

  \begin{figure}
\plotone{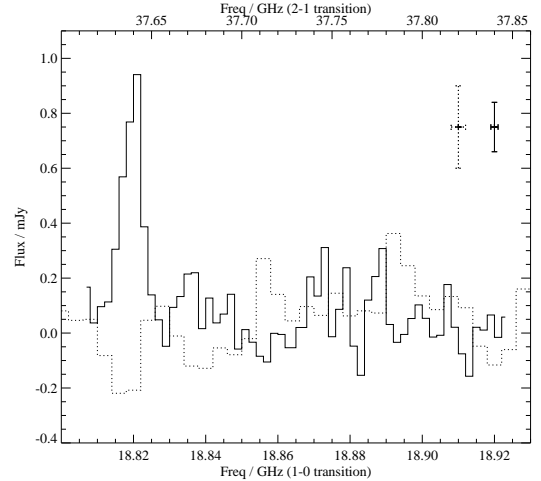}
\caption{The spectrum of our line-emitter source. Measurements at
18.8\,GHz (probing the CO(1-0) transition at $z=5.12$) are shown with
dotted lines, while measurements at 37.6\,GHz are shown with solid
lines. Typical error bars are shown for reference.
  \label{fig:spec}}
\end{figure}

This pilot survey at 37.7\,GHz probed only 3\% of the projected
spatial distribution and 35\% of the redshift range occupied by
clustered UV-luminous galaxies in our target field. Despite this, the
gas mass inferred from our single detection is comparable to the total
stellar mass in UV-luminous sources in the structure.  If this
detection is typical of the space density of molecular line emitters
in this field, then galaxies with unobscured star formation represent
at best a few percent of the baryons present \citep[consistent
with][]{2007astro.ph..1725V}. 

In the nomenclature of \citet{2005ARA&A..43..677S}, our detected
source is an Early Universe Molecular Emission Line Galaxy (EMG), a
broad category including radio galaxies, sub-mm galaxies and QSOs at
$z>2$.  It represents $10^6$\,M$_\odot$ of CO, formed within 1.1\,Gyr
of the Big Bang.  This yield is plausible from $3\times10^5$
population III stars, each of 100\,M$_\odot$, in as little as the
stellar lifetime \citep[10\,Myr,][]{2002ApJ...567..532H}. By contrast,
the yield of more conventional supernovae and the typical star
formation rates of UV-luminous sources in this field
($\sim$30\,M$_\odot$\,yr$^{-1}$), $10^9$ years would be required
\citep[see][]{2003Natur.424..406W} implying either early and
continuous star formation, or an early, intense starburst. A source
with unobscured star formation of 10\,M$_\odot$\,yr$^{-1}$ at $z=5.12$
would have been detected in our deep optical imaging, but observations
of dust emission in the rest-frame far-infrared will be necessary to
distinguish between a mature stellar population at the time of
observation and ongoing, but dust-obscured, star formation.

It is interesting to compare our detection with a similar observation
at $z=5.2$, also using the ATCA. \citet{2005ApJ...621L...1K} studied a
high mass system at this redshift, marked out by the presence of at
least six Lyman-$\alpha$ emitter galaxies and a powerful radio
galaxy. Based on UV-luminous stellar mass, this system should be
comparable in mass to our target region. \citet{2005ApJ...621L...1K}
detected CO emission in the 1-0 and 5-4 transitions, clearly
associated with the radio galaxy. While the inferred gas mass is
within a factor of four of our object, the velocity structure differs:
the emission line detected by \citet{2005ApJ...621L...1K} was twice as
broad as our line emitter, possibly due to momentum input from the
radio source.

We note that the CO luminosity and derived mass of our line-emitter
are similar to those of sub-millimeter galaxies (SMGs) at lower
redshifts \citep{2005MNRAS.359.1165G}. These are also often a
ssociated with galaxy overdensities at high redshift
\citep[e.g.][]{2003ApJ...583..551S}. While some SMGs with this mass
would have been detected in our deep {\em Spitzer}/IRAC imaging
\citep[based on their $K$-band magnitudes at $z=2.5$,][redshifted to
$z=5.12$]{2004ApJ...616...71S}, others would not.
CO emission is not a reliable star formation rate (SFR) indicator, but
is loosely correlated with far-infrared luminosity, and hence
SFR. Data on this relation at high redshift is sparse, but known EMGs
with comparable gas masses to our detected source have SFRs of
$\sim1000$\,M$_\odot$\,yr$^{-1}$. By contrast, the tight limit on
H$_2$ mass in the LBGs in our field would imply obscured star
formation rates of less than a few hundred solar masses per year, even
given dust properties similar to those of SMGs. Our detected source
likely represents a more extreme environment than that seen in LBGs at
the same redshift and could be a high redshift counterpart for the
intense starbursts in SMGs.

It is becoming increasingly clear that no one survey technique can
provide a complete picture of large scale structure at high
redshift. Overdensities of Lyman break galaxies are correlated with the
presence of Lyman-alpha emitters, radio galaxies, SMGs, DLAs
\citep{2006ApJ...652..994C}, and now molecular emission line galaxies
at the same redshift. Each component of the structure may account for
only a few percent of the baryonic mass and each presents
observational challenges. While LBGs and DLAs probe comparable mass
haloes at $z=3$ \citep{2006ApJ...652..994C}, molecular line emitters
may probe more massive and denser regions of the same 
structures.

Extensive extrapolation based on a single line from a single source is
risky. Observations of this source in other emission lines will be
required to confirm its redshift and luminosity. Such observations
will also begin to characterize the gas temperature and density, and
determine whether the CO to gas mass conversion factor for ULIRGs is
appropriate for low mass systems at early times. Additional
sources must be identified and characterized to determine whether our
initial detection is anomalous or typical of its
population. Ultimately, more extensive millimeter surveys, based on
existing deep datasets, are required to establish the molecular gas
content of large scale structures at high redshift, and to probe this
early stage of galaxy formation.

\section{Conclusions}
\label{sec:conclusions}

Our main conclusions can be summarized as follows:

(i) We have investigated the molecular gas content, as traced by
low-order transitions of CO, in a large scale structure identified in
Lyman break galaxies at $z=5.12$.

(ii) Neither of two known Lyman break galaxies in our survey area show
associated molecular line emission, suggesting that their molecular
gas content does not significantly exceed their stellar content.

(iii) We have identified a source with line emission at 37.642\,GHz,
which we identify as emission in the CO(2-1) transition at
$z=5.1245\pm0.0001$. This source is currently undetected at any other
wavelength.

(iv) The inferred gas mass in this single source is
M(H$_2$)$=(1.9\pm0.2) \times 10^{10}$ M$_\odot$, comparable to the
total stellar mass of UV-luminous sources in the same region.

(v) Further studies of UV-faint sources at high redshift are essential
to characterize such systems, and could make substantial progress
towards balancing the baryon budget at early times.

\acknowledgments

ERS acknowledges support from the UK STFC.  Based on data from ATCA
programs C1753 and C1821. The Australia Telescope Compact Array
is part of the Australia Telescope which is funded by the Commonwealth
of Australia for operation as a National Facility managed by CSIRO.
The authors thank Chris Carilli, Steve Longmore and Ron Ekers for
useful discussions. We also thank Maxim Voronkov and Robin Ward for
their vital assistance as Duty Astronomers at the ATCA.

{\it Facilities:} \facility{ATCA (12mm,7mm)}

\label{lastpage}

\end{document}